\documentclass[sigplan,screen,nonacm]{acmart}

\setcopyright{none}

\usepackage{mathpartir}
\usepackage{listings}

\definecolor{languagecolor}     {rgb}{0.5, 0.5, 0.5} 
\definecolor{ainfkeywordcolor}  {rgb}{0.7, 0.1, 0.1} 
\definecolor{leankeywordcolor}  {rgb}{0.0, 0.0, 1.0} 
\definecolor{samplekeywordcolor}{rgb}{0.6, 0.6, 0.1} 
\definecolor{tacticcolor}       {rgb}{0.0, 0.1, 0.6} 
\definecolor{commentcolor}      {rgb}{0.4, 0.4, 0.4} 
\definecolor{sortcolor}         {rgb}{0.1, 0.5, 0.1} 
\definecolor{attributecolor}    {rgb}{0.7, 0.1, 0.1} 

\definecolor{symbolcolor}      {rgb}{0.0, 0.1, 0.6} 
\definecolor{nosymbolcolor}    {rgb}{0.0, 0.0, 0.0} 
\newcommand{\symbolcol}{\color{symbolcolor}}

\newcommand{\bdiamond}[1][fill=black]{\tikz [x=1ex,y=1ex,line width=.1ex,line join=round, yshift=-0.285ex] \draw  [#1]  (0,.5) -- (.5,1) -- (1,.5) -- (.5,0) -- (0,.5) -- cycle;}%

\newcommand{\veryshortarrow}[1][3pt]{\mathrel{%
   \hbox{\rule[\dimexpr\fontdimen22\textfont2-.2pt\relax]{#1}{.4pt}}%
   \mkern-4mu\hbox{\usefont{U}{lasy}{m}{n}\symbol{41}}}}

\newcommand{\veryshortleftarrow}[1][3pt]{\mathrel{%
   \hbox{\usefont{U}{lasy}{m}{n}\symbol{40}}%
   \mkern-4mu\hbox{\rule[\dimexpr\fontdimen22\textfont2-.2pt\relax]{#1}{.4pt}}}}

\newcommand{\shortmapsto}{\mathrel{\mapstochar\mkern+2mu\veryshortarrow}}

\lstdefinelanguage{lean}{keywordstyle=[1]{\color{leankeywordcolor}},
  otherkeywords = {|, \#eval},
  morekeywords=[1]{
  axiom, match, with, end, if, then, else, let, in, class, instance,
  forall, exists, Pi, fun,
  example, extends, structure, deriving, variable, specialize, termination_by, pure, do, where, return, for, sum, by, rw, simp, rfl, next, obtain, have, cases, intro, subst, at, split,
  abbrev, def, inductive, |, \#eval,
  Prop, Type, Sort,
  SeqStRaw, List, AnfCont, Ctx, Env, Idx, Prim, Ren, Anf, Equation, Sat,
  SeqSt, Sim, Conc, AnfIsInv, AnfIsInvCont, SemiInvRaw, SemiInv, SigSyn, SigSem, Sig,
  Option, Fin, Nat,
  uniform, bernoulli, normal,
  },
  morekeywords=[2]{
  },
  keywordstyle=[3]{\bfseries\color{leankeywordcolor}}, morekeywords=[3]{theorem,},
  morecomment=[l][\itshape \color{commentcolor}]{--},
  basicstyle={\sffamily\normalsize}}
\lstdefinelanguage{scala}{keywordstyle=[1]{\color{scalakeywordcolor}},
  morekeywords=[1]{
  axiom, import, export, protected, private, public, override, infix, extension,
  val, var, def, class, object, package, trait,
  given, using, with, extends, deriving, implicit, summon,
  match, case,
  if, then, else, break, continue, return, try, catch, for, yield, do,
  macro,
  purify,
  }}
\lstdefinelanguage{rocq}{keywordstyle=[1]{\color{red}},
  morecomment=[s][\itshape \color{commentcolor}]{(*}{*)},
  morekeywords=[1]{
  Axiom,Definition,Check
  }}
\lstdefinelanguage{sample}{keywordstyle=[1]{\color{green!50!black}},
  morecomment=[l][\itshape \color{commentcolor}]{//},
  morekeywords=[1]{
  axiom,def,do,od,map,val,mk,pure,bind,extr,link,link2,link3
  }}
\newcommand{\languagetag}{}

\makeatletter
\lst@AddToHook{DeInit}{\languagetag}
\makeatother

\lstnewenvironment{leanlisting}[1][]{%
  \renewcommand{\languagetag}{}%
  \lstset{language=lean, basicstyle={\normalsize\sffamily}, #1}%
}{%
}
\lstnewenvironment{langlisting}[1][]{%
  \renewcommand{\languagetag}{\parbox[c][0pt]{\linewidth}{\raggedleft\scriptsize\color{languagecolor} \Lang}}%
  \lstset{language=lean, frame=lines, basicstyle={\sffamily\footnotesize}, keywordstyle=[1]{\color{ainfkeywordcolor}}, #1}%
}{%
}
\lstnewenvironment{ainflisting}[1][]{%
  \renewcommand{\languagetag}{\parbox[c][0pt]{\linewidth}{\raggedleft\scriptsize\color{languagecolor} \AINF}}%
  \lstset{language=lean, basicstyle={\ttfamily\footnotesize}, columns=[l]flexible, keywordstyle=[1]{\color{ainfkeywordcolor}}, #1}%
}{%
}
\lstnewenvironment{myequations}[1][]{%
  \renewcommand{\symbolcol}{\color{nosymbolcolor}}\noindent\adjustbox{center,vspace=\bigskipamount}\bgroup%
  \lstset{#1, frame=none}
}{%
  \egroup\renewcommand{\symbolcol}{\color{symbolcolor}}%
}

\lstMakeShortInline[language=lean,escapeinside={(*@}{@*)},basicstyle=\normalsize\sffamily,columns={[l]fullflexible}]!
\lstMakeShortInline[language=lean,basicstyle=\footnotesize\ttfamily,columns={[l]flexible},keywordstyle={[1]{\color{ainfkeywordcolor}}}]|

\begin{document}

\title[Extended Abstract: From Pattern Unification Towards Pattern Matching Unification]{
  Extended Abstract: From Pattern Unification \\ Towards Pattern Matching Unification}

\author{David Richter}
\orcid{0000-0002-8672-0265}
\affiliation{%
  \institution{Technische Universität Darmstadt}
  \city{Darmstadt}
  \country{Germany}
}

\author{Timon Böhler}
\orcid{0009-0002-9964-7367}
\affiliation{%
  \institution{Technische Universität Darmstadt}
  \city{Darmstadt}
  \country{Germany}
}

\begin{abstract}
We revisit the role of higher-order unification in dependently typed languages and identify a fundamental limitation of existing pattern-based fragments: their inability to synthesize functions defined by case analysis.
Even simple and ubiquitous constraints arising from type inference, particularly from use of induction principles, fall outside the expressive power of Miller patterns and their modern extensions.
We observe that such constraints naturally correspond to definitions by dependent pattern matching. Motivated by this correspondence, we propose integrating dependent pattern matching into the unification process. We present a prototype implementation of a small dependently typed language that collects delayed unification constraints and resolves them via a pattern matching compiler. Our approach successfully infers solutions that are rejected by current systems such as Rocq and Lean, suggesting a new direction for unification that unifies type inference and pattern matching compilation.
\end{abstract}

\maketitle

\section{Introduction}

Higher-order unification is an important part of type checking and type inference for dependently typed languages.
While higher-order unification is undecidable, dependently typed programming languages typically only make use of a fragment of full higher-order unification: Miller pattern unification.
Still, even expressive extensions of higher-order pattern unification fail to solve simple and ubiquitous equations.

Consider the constraint system:
\[
F\ \texttt{true} = 1
\qquad
F\ \texttt{false} = 2
\]

This problem is trivial from the perspective of functional programming: it asks for a function defined by case distinction on Booleans. The unique solution is:
\[
F := \lambda b.\ \texttt{match}\ b\ \texttt{with}\ 
\mid\ \texttt{true} \Rightarrow 1 
\mid\ \texttt{false} \Rightarrow 2
\]

However, this problem is not covered by Miller pattern unification~\cite{miller1991unification}, nor by the functions-as-constructors fragments (FCU)~\cite{libal2016functionsasconstructors} or the deterministic higher-order patterns (DHOP)~\cite{niederhauser2026dhop} extension, or extensions specific for dependently-typed languages~\cite{abel2011dynamicpattern}. All of these fragments fundamentally reject metavariable applications to terms without variables, and thus none of them can synthesize a function whose definition requires case analysis.

\section{Type Inference and Unification}

Such constraints are not even exotic; they are generated from type inference over eliminators.
Unification problems are generated for example from type checking function application, where the formal argument type must match the concrete argument type\footnote{this is already visible in the application rule for non-dependent functions}:
\[
\inferrule
{\Gamma \vdash f : A_1 \to B \quad \Gamma \vdash e : A_2 \quad A_1 = A_2}
{\Gamma \vdash f\ e : B}
\]

Furthermore, every time where implicit arguments appear non-linearly (more than once) in the type of a function,
this will lead to eventually having to solve a unification problem.
A particularly interesting situation is the use of induction principles.
Consider the induction principle generated for the Boolean type:
\[\begin{array}{ll}
\mathsf{Bool.ind} :& (P : \texttt{Bool} \to \texttt{Type}) \to \\
& P~\texttt{true} \to P~\texttt{false} \to \\
& (b : \texttt{Bool}) \to P~b
\end{array}
\]

During type checking of a term such as 
\[
\mathsf{Bool.ind}\ P\ 1\ \texttt{false}\ b
\]
the type checker will have to unify the expected and the actual type of the arguments.
The actual types are $1 : \texttt{Nat}$ and $\texttt{true} : \texttt{Bool}$.
The expected types are $1 : P\ \texttt{true}$ and $\texttt{false} : P\ \texttt{false}$.

Now suppose $P$ is a metavariable, i.e., the user omitted the motive of induction, hoping the proof assistant is able to infer it.
By unifying the expected and actual type, we obtain two constraints:
\[
P\ \texttt{true} = \mathbb{N}
\qquad
P\ \texttt{false} = \texttt{Bool}
\]

This is similar to the motivating problem from the introduction (except that the metavariable is now a type). Existing dependently typed languages such as Rocq and Lean are not capable of solving these unification constraints. Thus, we see that even simple uses of induction force us outside the pattern fragment.

The solution is of course:
\[
P := \lambda b.\ \texttt{match}\ b\ \texttt{with}
\mid\ \texttt{true} \Rightarrow \mathbb{N}
\mid\ \texttt{false} \Rightarrow \texttt{Bool}
\]

But how were we able to solve this so easily as humans?
Ignoring the theoretical underlay and considering these equations from a practical perspective as functional programmers,
we observe that they were in the familiar form of a function definition by pattern matching.

\section{Prototype}

We implemented a small dependently typed language in fewer than 3000 lines of Lean, designed to explore the interaction between unification and dependent pattern matching. The implementation includes intrinsically well-scoped syntax, untyped normalization by evaluation, and a type inference procedure with dynamic unification in the style of elaboration-based systems such as elab-zoo~\cite{kovacs2016elaborationzoo}. We additionally support inductive type declarations and a compilation procedure for dependent pattern matching.

The central idea of the prototype is to extend unification with a second phase that attempts to solve remaining constraints by synthesis via pattern matching. During type inference, we collect higher-order unification constraints that fall outside the supported pattern fragment. Instead of rejecting them, we suspend these constraints and record them as delayed problems, consisting of equations between a metavariable applied to arguments and a right hand side, as is usual in dynamic unification.

After type inference, the delayed problems are passed to the dependent pattern matching compiler. Intuitively, each problem is reinterpreted as a definition to be synthesized: the metavariable becomes a function symbol, its arguments become pattern variables, and the constraint specifies the desired output. The pattern matching compiler then attempts to construct a definition by case analysis on the arguments.

Operationally, this amounts to reusing the existing pattern matching compilation machinery as a solver for a restricted class of higher-order constraints. Successful compilation yields a definition that is fed back into the unification context, discharging the original constraint. If compilation fails, the constraint remains unsolved and is reported as an error.

This approach is particularly interesting for constraints arising from eliminators and induction principles, where the missing term is naturally defined by case analysis. In Figure~\ref{fig:ex}, we return to the example from the previous section but give the actual code that produces this unification problem in three languages: This work, Rocq and Lean. For simplicity, we assume the induction principle as an axiom. Then, using the induction principle with different types in the branches induces a non-pattern unification problem which is rejected by Rocq and Lean, but is successfully resolved in our prototype via pattern matching synthesis.

\begin{figure}
\raggedright
\textbf{This Work} (pseudocode):
\begin{lstlisting}[language=sample]
axiom ind P b : P true → P false → P b
def foo b := ind _ b 1 false
// solution ?P _₀ :=
//   match _₀ with | false => Bool | true => Nat
\end{lstlisting}

\medskip

\textbf{Rocq:}
\begin{lstlisting}[language=rocq]
Axiom ind : forall (P : bool -> Type) (b : bool),
  P true -> P false -> P b.
Check fun b => ind _ b 1 false.
(* Unable to unify "bool" with "?P false"
  (cannot satisfy constraint "?P false" == "bool"). *)
\end{lstlisting}

\medskip

\textbf{Lean:}
\begin{lstlisting}[language=lean]
axiom ind : ∀ (P : Bool → Type) (b : Bool),
  P true → P false → P b
def foo b := ind _ b 1 false
-- Application type mismatch:
-- The argument                          false
-- has type                                  Bool
-- but is expected to have type   ?m.6 b false
\end{lstlisting}
\caption{Pattern Matching Unification Litmus Test}
\label{fig:ex}
\end{figure}

\section{Limitations and Future Work}

Our approach inherits both the strengths and the limitations of dependent pattern matching. In particular, it can only solve unification problems that can be expressed as definitions by dependent pattern matching. While this already covers important cases such as motives of induction principles, the precise boundary of the solvable fragment deserves further study. Beyond simple examples such as Booleans, it would be valuable to characterize systematically which classes of constraints can be handled, including those arising from indexed families and GADTs.

Our current prototype supports dependent pattern matching over inductive families (including equality types), but omits recursion. Extending the approach to recursive definitions is a natural next step. Standard higher-order unification enforces acyclicity of metavariable solutions, whereas functional programs routinely rely on structurally recursive definitions. This suggests replacing the acyclicity check with a structural recursion check, as already used in systems such as Rocq, Agda, and Lean. Such an extension could enable the synthesis of recursive functions as solutions for metavariabels during unification.

A further direction is the treatment of mutually recursive definitions. If recursive metavariables can be solved via pattern matching, it is natural to ask whether mutually recursive metavariables can be inferred analogously, potentially increasing the expressiveness of type inference for programs involving mutually defined functions.

From an implementation perspective, our current system produces case trees as solutions to unification problems, but does not yet translate them into core language terms. Completing this compilation step and integrating it with existing elaboration pipelines is necessary for a practical system.

More broadly, our work suggests a tighter connection between unification and elimination: inferring elimination principles appears to require unification procedures that can themselves perform elimination. Making this correspondence precise, and understanding its implications for the design of type inference algorithms, remains an interesting direction for future research.

\section{Conclusion}

Besides the usual use case of higher-order unification for type checking and type inference, modern dependently typed languages often have a second use for higher-order unification, namely to compile dependent pattern matching~\cite{cockx2018elaborating,sozeau2019equations}.
As such, previous work has considered the question: \emph{Can we use unification for dependent pattern matching}.
The question we ask in this paper now goes in the other direction: \emph{Can we use dependent pattern matching for unification?}

We explore integrating dependent pattern matching into the unification algorithm for type checking and inference, providing a more expressive mechanism for solving metavariables, particularly for motives of induction principles. This suggests a unification of unification and pattern matching.

\bibliographystyle{plain}
\bibliography{bibliography}

\end{document}